\documentclass[12pt]{article}

\input epsf
\def\beq{\begin{eqnarray}}
\def\eeq{\end{eqnarray}}

\def\L*{{\cal L}_*}
\def\lsim{\mathrel{\rlap{\lower3pt\hbox{\hskip0pt$\sim$}}
     \raise1pt\hbox{$<$}}}         
\def\gsim{\mathrel{\rlap{\lower4pt\hbox{\hskip1pt$\sim$}}
     \raise1pt\hbox{$>$}}}         

\usepackage{amsmath}
\usepackage{amsfonts}
\usepackage{amssymb}

\begin{document}

\begin{titlepage}

\thispagestyle{empty}

\begin{flushright}
{NYU-TH-07/12/24}
\end{flushright}
\vskip 0.9cm

\centerline{\Large \bf Electrodynamic Metanuclei}                    

\vskip 0.7cm
\centerline{\large Gregory Gabadadze and  Rachel A. Rosen}
\vskip 0.3cm
\centerline{\em Center for Cosmology and Particle Physics}
\centerline{\em Department of Physics, New York University, New York, 
NY, 10003, USA}

\vskip 1.9cm

\begin{abstract}

A relativistic system of electrically charged fermions and oppositely 
charged massive scalars with no self-interactions, is 
argued to have a long-lived collective state 
with a net charge.  The charge is residing near
the surface of the  spherically-symmetric  state, while the  
interior consists of the condensed scalars, that are neutralized 
by the fermions. The metastability is achieved by 
competition of the negative pressure of the scalar condensate,
against the positive pressure, mainly due to the fermions.
We consider such metanuclei  made of  
helium-4 nuclei and electrons, below nuclear but above 
atomic densities.  Typical  metanuclei represent
charged balls of the atomic size, colossal mass, electric 
charge and excess energy.  Unlike an ordinary nucleus,  
the charge of a metanucleus scales proportionately  to its radius. 
The quantum mechanical decay through tunneling, and 
vacuum instability  via pair-creation, are both suppressed for 
large values of the electric charge. Similar states
could also  be composed of other charged (pseudo)scalars, 
such as the pions, scalar supersymmetric partners, or in general, 
spin-0 states of new physics.

\end{abstract}

\vspace{3cm}

\end{titlepage}

\newpage

\section{Introduction and summary}

The purpose of this work is to show that 
in a relativistic system of $N+Q$ fermions, each of charge $g$, 
and $N$ oppositely charged massive scalars, with no non-linear 
self-interactions,  there may exist a metastable long-lived 
spherically symmetric ball  with the following identity: 
The excess charge $gQ$ is residing on a surface of the 
ball, while in its neutral interior there are $N$ 
condensed scalars, that act collectively as a macroscopic state of a 
large occupation number, and  also the $N$  fermions playing the 
role of spectators that neutralize the 
bulk scalar charge\footnote{We should also make sure that the fermions 
and scalars  don't  form neutral atoms, this could be arranged 
by increasing density and/or temperature of the system (see below).}.
The radius of the ball $R$ scales linearly with the charge $gQ$, and 
the  electric field near its surface, $gQ/(4\pi R^2)$, 
decreases with increasing  charge, in the regime 
of applicability of our arguments, $N^{1/3}\ll g^2 Q\ll gN^{2/3}$.

The physical reason for (meta)stability of such a ball 
is that the scalar condensate gives an attractive 
negative pressure which balances against the repulsive pressure, 
mainly due to the positive energy of the fermions.  
As a result, the energy functional has a minimum around the 
point of balance. The total energy
stored in the ball is greater than the energy of 
$N$ neutral scalar-fermion atoms plus free $Q$ fermions.
Hence, the minimum of the energy functional is only a 
local one -- the condensate ball can decay into the 
atomic state by tunneling. However, for large values of 
the charge, both the tunneling and  the vacuum 
instability through quantum pair-creation,  
are suppressed.

The spectrum of small perturbations above the scalar condensate in the 
bulk of the  ball has a mass 
gap that equals to $2m_H$, where $m_H$ is the mass of the scalar. 
Moreover, the photon becomes massive, 
with its Compton wavelength smaller than the size of the 
condensate ball \cite {GGRR}. We refer to these balls as metanuclei.

Although the easiest way to understand the (meta)stability of the 
metanuclei is in terms of the balance between the positive (outward) 
pressure of the bulk fermions,  and  negative  pressure of the charged 
condensate, it is  nevertheless useful to describe  this in terms of the 
electrostatic interactions too.  In the bulk of the 
metanucleus, there is a screened interaction  
between the negatively charged fermions and positively charged condensate. 
Furthermore, there is an attractive interaction
between the charged condensate and surface fermions, 
and repulsion between  the bulk and surface fermions, as well as 
between the surface fermions themselves.

In an analogous problem with positively 
and negatively charged classical particles in the bulk, and negatively 
charged particles on the surface of radius $R$, 
there would be an exact cancellation 
between  the  interactions of  the positively  and negatively 
charged bulk particles with the negatively charged surface particles;
hence, such a surface would not be stable.

However, this is not the case in our problem.
Charged condensate in the bulk differs from just a collection 
of classical charged particles.  The former is a  relativistic 
substance  and its attractive interaction with the 
surface fermions scales differently with $R$  \footnote{A field they 
reason for this is that  the expression for the current density of fermions 
does not depend on the gauge field, while that for  a relativistic 
scalar manifestly depends on the gauge field via the covariant derivative.
Note that in the Thomas-Fermi (or any other mean field) approach
the effective fermion number density would depend on the value of the 
gauge potential, however, the latter would still be different from 
the dependence of the number density for the scalar.}.
Moreover, in the regime when  
$N^{1/3}\ll g^2 Q\ll gN^{2/3}$, the bulk-surface interactions,
which end up being of order $NQ$,  dominate over the net bulk-bulk
(at most $\sim N^{4/3}$), 
and the surface-surface (at most  $\sim Q^2$) interactions. 
Precisely it is in this regime that  we find the local minimum 
of the energy functional. 

In spite of a seeming mutual repulsion  of the $Q$ fermions 
at the surface, it is energetically unfavorable for them to escape, 
as we will show in Section 4. The presence 
of the $Q$ fermions  on the surface, is the very reason for
the existence of a constant gauge invariant potential in the bulk,
which is responsible for keeping the bulk scalars 
in the  condensate state.  Hence, removal of surface 
fermions becomes energetically expedient, since it causes 
changes in the bulk energy.

In a certain respect, the metanuclei  resemble properties of 
a metallic ball with an excess charge\footnote{Except that, 
we do not expect the charged condensate to form a 
crystalline structure, since interactions in its bulk  
are screened, see discussions in \cite{GGRR}.}.
In metals, the excess 
charge resides on  the surface because it's energetically 
favorable to maintain zero electric field in the bulk.  
The surface charge, when it's not too large, in spite of its mutual 
repulsion, is not escaping the metallic ball 
(see more in Section 4).  

The above construction seems generic. It could be applicable to systems 
that are described by a relativistic Abelian gauge theory. The 
scalars and fermions  could  be fundamental particles or 
composite states.

One application is to the system of  $2N+Q$ electrons, $e^{-}$, 
and $N$ helium-4  nuclei, $He^{++}$. We consider this system 
below the nuclear but above atomic densities, so that the nuclear 
effects are negligible, while the atoms are dissolved.   We show that
long-lived  charged metanuclei, made of the 
electrons and condensed $He^{++}$ states, may exist. 
These  metanuclei are truly colossal -- of the size of the 
Hydrogen atom  or even greater -- and carry enormous charge and 
excess energy. Such giant nuclei cannot form neutral atoms, 
making their survival in the Universe potentially difficult.

The metanuclei could  also be ``made of'' 
other existing particles, such as pions,  
or still hypothetical particles, such as 
sleptons or squarks, if captured in 
the condensate before they decayed.

The organization of the paper is as follows: in Section 2 we zoom onto the 
interior of the condensate ball assuming that it fills entire  space
(this is a good approximation for the ball as long as its size is much 
greater  than the width of its boundary  region; we'll justify this 
assumption for the metanuclei in Section 3). We briefly summarize 
the results of \cite {GGRR} on condensation of charged scalars. 
In Section 3 we discuss the surface-bulk connection and show that 
the condensate ball  is a (local) minimum of the energy 
functional. In Section 4 we discuss energetics of the charged 
condensate balls and  study why a few possible decay channels of 
these balls can be suppressed. In Section 5 we give some examples.

The metanuclei resemble a non-topological soliton (see, e.g., 
\cite {Rosen,Sirlin,Coleman}) of a $Q$-ball type 
\cite {Rosen,Coleman,Kusenko} with a local charge 
(charged Q-balls) \cite {Lee} -- especially the charged 
Q-balls with fermions \cite {Tetradis}. However, there are 
important differences: the $Q$-balls require a special form of 
the  non-linear potential for the scalar \cite {Rosen,Coleman,Lee}, 
while in our  case the non-linear scalar self-potential plays no 
role -- it is the scalar mass term and its 
interactions with the gauge field that are  crucial.
$Q$-balls could form  due to a global minimum
of the energy functional, while  the 
metanuclei would form due to a local minimum.

\section{Dynamics in the interior (bulk) of the ball}

We start by considering a simple model of charged scalars 
and oppositely charged fermions at zero temperature.  
The classical Lagrangian contains a gauge field $A_\mu$, 
a charged scalar field $\phi$ with mass $m_H$, and fermions 
$\Psi^+,\Psi$ with mass $m_J$
\beq
{\cal{L}} = -\tfrac{1}{4}F_{\mu\nu}^2 + 
\vert D_{\mu} \phi \vert^2 - m_H^2 \phi^{\ast} \phi +
{\bar \Psi}(i\gamma^\mu D_\mu  -m_J) \Psi \,.
\label{lagr0}
\eeq
The covariant derivatives in (\ref {lagr0}) are defined 
as  $\partial_\mu -ig A_\mu$ for the scalars and  
$\partial_\mu +ig A_\mu $ for the fermions. Although, for 
simplicity we have assumed that the scalar and 
fermion charges are equal, $g \equiv g_{\psi} = -g_{\phi}$, 
our results apply to a general case\footnote{The conditions under which 
the other possible interactions in the Lagrangian  (\ref {lagr0}) 
won't affect our conclusions were discussed in Ref. \cite {GGRR}.
For instance, there is a wide range of the parameter space where
the possible scalar quartic coupling term is insignificant for our 
discussions.}.

We introduce the following notations for the scalar, fermion, 
and gauge fields: $\phi = \tfrac{1}{\sqrt{2}} \sigma\, e^{i \alpha}$, 
$\Psi = \psi e^{-i \beta}$,
${B_{\mu}} \equiv A_{\mu} - \tfrac{1}{g}\partial_{\mu} 
\alpha$, and $\gamma \equiv
\alpha - \beta$. In terms of the {\it gauge invariant} variables 
$\sigma$, $\psi$, ${B}_\mu$ and $\gamma$, the Lagrangian takes the form
\beq
{\cal{L}}=-\tfrac{1}{4}F_{\mu\nu}^2 + 
\tfrac{1}{2}(\partial_{\mu}\sigma)^2+
\tfrac{1}{2} g^2{B}_\mu^2 \sigma^2- \tfrac{1}{2}m_H^2 
\sigma^2 + {\bar \psi}(i\gamma^\mu D_\mu  -m_J) \psi -(\partial_\mu\gamma)
{{\bar \psi}\gamma^\mu  \psi},
\label{lagr}
\eeq
where now $F_{\mu\nu}$ and $D_\mu$  are the field-strength and covariant 
derivative for ${B}_\mu$, respectively.  
The key point for our discussions is that the third term 
in the Lagrangian (\ref {lagr}) gives rise to a tachyonic mass
for the scalar $\sigma$ if  the field 
$gB_{0}$ acquires a vacuum expectation value \cite{Linde,Kapusta}. 
Moreover, when $\langle gB_{0}\rangle =m_H$, the scalar field condenses. 
Note that we retain the (last) total-derivative term in 
(\ref {lagr}) as it will be important in our considerations. 

To reach  the condensate point, following  Ref. \cite{GGRR}, 
we assume that temperature is low-enough that the fermions 
form a degenerate system with overlapping de Broglie wavelengths,
so that  the fermions can be averaged over. Hence, 
we consider a system with a uniform background of fermions: 
$J_\mu \equiv {\bar \psi}\gamma_\mu \psi = J_0 \delta _{\mu 0}$. 
The equations of motion derived from (\ref{lagr}) are:
\beq
-\partial^\mu F_{\mu\nu} = g^2 B_\nu \sigma^2 -g J_\nu \,,~~~~~
\square \sigma = g^2 B_\mu^2 \sigma - m_H^2 \sigma \,.
\label{Name}
\eeq
The theory admits a static solution with constant  
$B_0$, $\sigma$: 
\beq
\langle g B_0 \rangle = m_H \,,~~~~~~~
\langle \sigma \rangle  = \sqrt{\frac{J_0}{m_H}} \,.
\label{b0}
\eeq
The quantity $-\langle g B_0 + \dot{\gamma} \rangle$ 
acts as a dynamically induced  chemical potential for the fermions, 
implying $ \epsilon^\prime _F =- \langle g B_0 + \dot{\gamma} \rangle$,
where $\epsilon^\prime _F$ is the Fermi energy (unconventionally 
normalized, to include $B_0$ for simplicity).
By this  the negative value of  the phase, 
$\dot{\gamma} \equiv \partial_0 \gamma$, gets fixed.

For the scalars, it is the quantity $\langle g B_0 \rangle= m_H$ that 
acts as an effective chemical potential\footnote{One  could have 
also introduced a 
chemical potential $\mu_s$ for the scalars  by adding the terms 
$+\mu_s (-g B_0 \sigma^2)+ \tfrac{1}{2}\mu_s^2 \sigma^2$ to 
(\ref{lagr}). However, we can absorb these terms into a 
redefinition of $B_0$: $B_0' = B_0-\tfrac{1}{g} \mu_s$.}.  

The bulk of the condensate is electrically neutral
due to the compensation between the fermion  and scalar 
charge densities: $gJ_0-g^2B_0\sigma^2=0$. However, a nonzero $gB_0$ 
implies an uncompensated charge on a surface enclosing the condensate 
\cite{GGRR}. The spectrum of small perturbations above the condensate
is composed of a scalar of mass $m_s=2m_H$, and a  
photon that has acquired the mass $m_g=g \sqrt{J_0/m_H}$ 
(see, Ref. \cite{GGRR} for details). 

The purpose of the present work is to show that such objects, 
with the surface charge and condensate bulk, can be long-lived. 
For notational simplicity, from now on we will be dropping 
the brackets, $\langle \cdot \rangle $, denoting the condensates.

\section{Surface-bulk connection}

We now consider a spherically symmetric system of a 
finite radius $R$ and look for a (meta)stable solution.  We include the 
dynamics of the fermions in our considerations.  
The Hamiltonian derived from (\ref{lagr}) is
\beq
{\cal{H}}&=&{\cal{H}}_\psi+\frac{1}{4} F^2_{ij}+\frac{1}{2}\pi_j^2
+ {B}_0 \left(\partial_j \pi_j+g J_0-{1\over 2} g^2 B_0 \sigma^2 \right) 
+\frac{1}{2} 
P_\sigma^2 \nonumber \\
&+& \frac{1}{2} \left(\partial_j \sigma \right)^2
+\frac{1}{2} g^2 {B}_j^2 \sigma^2
+\frac{1}{2} m_H^2 \sigma^2 \,,
\label{ham0}
\eeq
where $\pi_j \equiv -F_{0j}$, $P_\sigma = \dot{\sigma}$, are 
the canonical momenta  for the ${B}_j$ and $\sigma$ 
fields respectively, and ${\cal{H}}_\psi \equiv i{\bar \psi} \gamma_j 
(\partial_j+ig {B}_j +i \partial_j \gamma) \psi +
m_J  {\bar \psi} \psi$, denotes the 
Hamiltonian density of the fermions\footnote{The quantity $\gamma$ is not 
a dynamical field. We will find  a nonzero 
$\partial_0 \gamma$ on the solution. The latter can be though of  
as the chemical potential for the fermions.
Upon  canonically transformation  from the Lagrangian density 
(\ref {lagr}) to the Hamiltonian density  (\ref {ham0}) the 
total derivative term with $\gamma$ disappears. 
As a result,  the expression  (\ref {ham0})
corresponds to a Hamiltonian density ${\cal H}$, and not to the thermodynamic 
potential density,  ${\cal H}^\prime \equiv {\cal H}-\mu J_0$,  
of the Grand Canonical partition function.  The energy that is being 
minimized is determined by ${\cal H}$, while the equations of motion are 
those obtained from ${\cal H}^\prime $, i.e., one has to add 
to (\ref {ham0}) the chemical potential  term (expressed via $\dot \gamma$) 
to recover the correct Lagrangian equations of motion.}.

As mentioned above, a nonzero $g B_0 +\dot{\gamma}$ acts as an 
effective chemical potential for the fermions in the ball: 
\beq
\epsilon^\prime_F \equiv \sqrt{(3 \pi^2 J_0)^{2/3}+m_J^2} =- ( 
g B_0 +\dot{\gamma}) \, .
\label{sigmaeq}
\eeq
This defines the value of $\dot{\gamma}$ in the bulk to be 
$\dot{\gamma} =-(\epsilon^\prime_F +m_H)$, and we regard to $B_0$ as 
a part of the total chemical potential for the fermions 
\footnote{One could apply 
the Thomas-Fermi (TF) approach to the system of fermions. 
In ordinary electrodynamics 
this would lead to a selfconsistent  non-linear equation for the 
gauge potential.  Here, the TF  equation  contains  the phase of 
the charged  scalar (or, equivalently,  its gauge-invariant 
combination with the phase of the fermion, $\gamma$),  
which in our notations appears in the fermion equation 
as an effective  chemical potential.
The TF equation just determines that phase. The result for 
the fermions in the bulk of the ball is automatically 
accounted for by our eq. (\ref {sigmaeq}). We will use 
below the TF approach for  the fermions near the surface,
see eq. (\ref {TF}).}.

As $\dot{{B_0}}$ does not appear in the Lagrangian, 
the equation of motion for ${B}_0$ 
gives us Gauss's law:  
\beq
-\nabla^2 B_0 +\partial_0 \partial_j B_j \, = \,g J_0-g^2 B_0 \sigma^2 \, 
\equiv \, g J_0^{\rm{total}}  \, .
\label{gauss}
\eeq
\noindent Equation (\ref{gauss}) has two important implications for the value 
of the fields in the bulk.  The first is that in the bulk of the condensate 
where gradients are zero we have
\beq
gB_0 = \frac{J_0}{\sigma^2} \, .
\label{gb0}
\eeq
Secondly, equation (\ref{gauss}) determines the value of $B_0$ in the 
bulk in terms of the conserved charge $Q$ and the radius $R$.  
Taking $\partial_0 \partial_j B_j$ to be zero everywhere, we 
solve equation (\ref{gauss}):
\beq
B_0(r) = \left\{
	\begin{array}{ll}
	\frac{gQ}{4 \pi R}\, & {\rm{for }} ~ r \leq R\, , \\  \\
	\frac{gQ}{4 \pi r}\, & {\rm{for }} ~ r >  R\, ,
	\end{array}  \right.
\label{bcharge}
\eeq
where $Q \equiv \int{d^3r J_0^{\rm{total}}}$.   
We have set $B_0 \rightarrow 0$ as $r \rightarrow \infty$, since  
$B_0$ is a gauge invariant variable, 
and a nonzero $B_0$ in the vacuum (i.e., far away from the condensate ball) 
would imply a different  spectrum  of the theory - 
a different mass for $\sigma$ and Lorentz violating 
interactions of  $\sigma$ with  the gauge field.

Therefore, we can use (\ref{gauss}) to integrate out $B_0$ 
from the Hamiltonian, which becomes
\beq
{\cal{H}} = {\cal{H}}_\psi+ \frac{1}{2}
\frac{J_s^2}{\sigma^2}+
\frac{1}{2} m_H^2 \sigma^2 +{\cal{H}}_{\rm{surface}} \, ,
\label{ham1}
\eeq
\noindent
where $J_s=J_0+g^{-1}\partial_j\pi_j$ is the scalar charge density, 
and ${\cal{H}}_{\rm{surface}}$ refers to all surface 
and gradient terms.  In order not to select a preferred 
direction we set $B_j$, $J_j$ to zero. 

Note that one can apply the scaling arguments  to the expression 
for the energy functional that is 
obtained by integrating (\ref {ham1}) w.r.t. $d^3x$. 
The scaling  considerations give an opposite dependence on the scale parameter 
of the second and the third terms on the r.h.s. -- one is a 
decreasing  function of the scale 
parameter while the other one  is an increasing function. 
This suggests that there should exist  at least a local 
minimum of the energy functional due to the competition 
between these two terms, when they dominate over the others.

At this point we have used every equation of motion except for the 
equation  of $\sigma$ and Gauss's law. 
For fixed scalar charge density $J_s$,  the 
second and third terms on the r.h.s. of (\ref {ham1}) could be 
thought of as an effective potential for the 
$\sigma$ field, in the regime where the 
field does not change  significantly.
In that regime, the above potential has a minimum.  
We vary (\ref{ham1}) with respect to $\sigma$, 
ignoring all the gradient terms, and find, 
$\sigma = (J_0/m_H)^{1/2}$. Using this in equation 
(\ref{gb0}) we find  that $g B_0 = m_H$. This is consistent with the 
solution of the previous section. Moreover, from (\ref{bcharge})
we deduce
\beq
R_c={\alpha_g Q\over m_H}, ~~{\rm where}~~\alpha_g \equiv 
\frac{g^2 }{4 \pi}\,.
\label{Rc0}
\eeq
Thus, for a given $Q$, the radius of a ball of condensate 
is completely determined.\\

Furthermore, we wish to show that the radius (\ref {Rc0}) 
minimizes the energy of the condensate ball as a functions of $R$, 
in agreement with the scaling arguments. The formalism of the previous 
paragraph does not allow us to do so as $R$ is fixed on the solution.  
Instead we relax our enforcement of the equation for $\sigma$, and vary w.r.t. 
$R$.  Our logic is as  follows:  In the bulk $B_0 = gQ/(4 \pi R)$.  
In addition to the charge $Q$ being conserved, the total number of scalars 
$N_s$ is also conserved:
\beq
N_s = \int{d^3r g B_0 \sigma^2} \, .
\label{n_s}
\eeq
Then, from the scaling of $B_0$ in the bulk, $B_0 = gQ/(4 \pi R)$,  
it follows that  $\int{d^3r \sigma^2} \sim R$.  
Using these scalings in (\ref{ham1}),
the total energy 
dependence on $R$ can be read from:
\beq
E=E_\psi+\frac{N}{2} \left(\frac{\alpha_g Q}{ R}+
\frac{ m_H^2 R}{\alpha_g Q}\right) + E_{\rm{surface}} \,,
\label{energy1}
\eeq
where $N \equiv \int{d^3r J_0}$ is the total number of fermions 
and $N \simeq N_s$ as long as $Q \ll N$. The first term on the r.h.s. 
of equation (\ref{energy1}) is the energy 
of the free fermions which, due to their degeneracy pressure,  
tend  to expand the ball of condensate. The first term in the 
parenthesis comes from the scalar-gauge and fermion-gauge field 
interaction terms and also provides positive pressure.   
The term $E_{\rm{surface}}$ contains the non-relativistic part of the 
energy due to the surface charge, which 
works to expand the ball as well.  It is only the second 
term in the parenthesis in (\ref{energy1}), however,
that provides the negative pressure and wants to contract the ball.
Because this term contains $\sim \int{d^3r \sigma^2}$,  it scales as $R$.  
We can use this  negative pressure to stabilize the ball against 
the other terms. 

We chose to consider solutions where 
the repulsive term $\sim NQ/R$, and the attractive 
negative pressure term $\sim m_H^2 N R/Q$ are dominant\footnote{
We could also stabilize the 
negative pressure term against any other positive pressure terms 
in the above expression, e.g., against the fermion degeneracy pressure term.  
However, the solutions obtained by stabilizing against 
$E_\psi$ or $E_{\rm{surface}}$ do not recover  the infinite volume 
solution in the bulk of the ball. Although these solutions may 
well exist,  their properties, such as a spectrum of small 
perturbations, would be different.}. In the limit that the 
fermions are relativistic, this is true when  $\alpha_g Q \gg N^{1/3}$.  
For non-relativistic fermions, the bound is 
$\alpha_g Q \gg (m_H/m_J)^{1/2} N^{1/3}$.  
In either case, the critical radius is in agreement with (\ref{Rc0}),
obtained previously from the variation w.r.t. $\sigma$.

The exact static solution of the equations of motion (\ref {Name}) 
is hard to obtain. In Ref. \cite {GGRR} we found an approximate solution 
in the interior and exterior of the condensate ball. For generic values 
of the parameters, the obtained solutions are valid everywhere 
except in a very narrow region near the boundary of the ball, 
where our approximations break 
down\footnote{The solutions are  valid near the boundary as well  
only for particular values of the parameters.}.   Nevertheless, 
we matched the asymptotic solutions and their derivatives 
across the surface,  demonstrating that with the asymptotic 
boundary conditions that we used, there are enough integration 
constants for the matching to be possible. The matching gave a 
relation between  the critical radius $R_c$ and charge $gQ$ 
which closely approximates (\ref {Rc0}).

In our derivations of the solution 
in Ref. \cite {GGRR}, and in its use here,  we assumed that there is a bulk 
region in which the  fermion number density is homogeneous, and that 
the size of this region is much greater than the surface width 
(i.e., we used the thin-wall approximation).  
Having made this assumption,  we solved  the coupled classical  
equations of motion and found that the width of the surface 
is determined  by the scale $(m_Hm_g)^{-1/2}$. 
The latter happens  to be  much smaller  then the size of the bulk 
region, $R_c\simeq \alpha_g Q/m_H$, as long as $NQ \gg \alpha_g^{-1}$, 
which is the case here. 

To complete the check of the thin-wall assumption the 
matching of the fermionic   component should also be considered. 
For this, we use the Thomas-Fermi 
approach. Since the fermions  couple to the net potential  
$B_0+{\dot \gamma}$,  the TF equation in our case takes the form:
\beq
-\nabla^2B_0 + g^2 B_0 \sigma^2 = { g\over 3\pi^2}
\left ( (gB_0 + {\dot \gamma})^2 -m_J^2 \right )^{3/2}\,.
\label{TF}
\eeq
The above equation can be solved  inside and outside of the radius $R_c$.
The interior solution coincides with that of  the previous section. 
In the exterior,  we obtain 
\beq
{\dot \gamma}_{\rm ext} \simeq -{\alpha_g Q\over r}
- \sqrt{(3\pi^2 g B^{\rm  ext}_0 \sigma_{\rm  ext}^2)^{2/3}+  m_J^2}\,,
\label{gammaext}
\eeq 
where  $B^{\rm  ext}_0$ and 
$\sigma_{\rm  ext}$ are the exterior solutions  of \cite {GGRR}.
The interior and exterior solutions
for ${\dot \gamma}$ match at $r \simeq R_c$, as they should. 
This is enough for the complete matching,  since in our formalism 
$\gamma$ has no second derivative term in the action. 
As an outcome of the above considerations, and 
using the  fact  that the $\sigma$ field  vanishes  
exponentially for $(r-R_c)\gg m_H^{-1}$, we find that 
${\dot \gamma} \simeq -(\alpha_g Q/r)-m_J $, 
away from the ball.  

As was pointed out above, the total effective potential for the 
fermions (the potential plus the chemical potential) 
is, $B_0+{\dot \gamma}$. Its value is a negative 
constant in the interior, $-\sqrt{(3\pi^2J_0)^{2/3}+m_J^2}$, 
while in the exterior it asymptotes to $-m_J$. Therefore, the reletivistic 
fermions appear as if they're trapped in  a potential well of the depth 
$\sim J_0^{1/3}$ and width $\sim \alpha_g Q/m_H$. This should 
certainly be so for an equilibrium state.

\section{Energetics}

In order to determine whether the condensate ball can be absolutely 
stable or not we  should compare its energy  with the total 
energy of $N$ neutral atoms formed by the scalars and fermions, 
and $Q$ free fermions. This energy is: 
\beq
E_a= (m_H - E_b)N + m_JQ\,,
\label{aenergy}
\eeq 
where the binding energy is determined by 
$E_b \simeq (\alpha_g)^2 m_Jm_H/2(m_J+m_H)$.

The energy of the condensate ball can be  
calculated from (\ref {energy1}) using 
(\ref{Rc0}):
\beq
E_c=E_\psi+m_HN+ E_{\rm{surface}} \,.
\label{energy2}
\eeq
The latter would always exceed (\ref {aenergy}). 
However, even when  $E_c$ is greater than $E_a$ the condensate ball 
could be a long-lived as it represents a local minimum of the 
energy functional. In this case, it will be classically stable, however, 
would be able to decay through tunneling.
To estimate the probability of the tunneling
one could use an analog quantum mechanical decay rate,
\beq
\Gamma \propto {\rm exp}\left ( -\int^{R_b}_{R_c}{dR\,(E(R)-E_c)} 
\right),
\label{drate}
\eeq
where $R_b$ is an initial radius of the ball after the tunneling. 
The ball could tunnel, while radiating away energy, directly 
into the size   $R_b \sim R_a \equiv N^{1/3}/(\alpha_g\, m_J)$ 
that would  allow  the state of  $N$ neutral atoms and 
$Q$ free charges to form.
However, $R_a$ is much greater than $R_c$ according to our construction, 
and such a process would be highly suppressed.  Instead, the ball 
could first tunnel into a state of a radius smaller than $R_a$ 
but greater that $R_c$, and then expand  toward the state with neutral atoms. 
An  estimate of the tunneling  
rate  for the latter process could be obtained by assuming 
that $(R_b-R_c)\sim R_c$ and $(E(R_b)-E_c)\sim E_c$, 
this being justified when $m_H$ is the heaviest mass scale, and implies that 
individual particles have to overcome at least a potential barrier 
with the hight of order $\sim m_H$, and widths of  order $\sim R_c$.
Then, using the expressions $E_c \simeq m_HN$ and 
$R_c =\alpha_gQ/m_H$, we get the following scaling for 
the decay rate, $\Gamma \propto \exp{\left(-k\alpha_g N Q \right)}$,
where $k$ is some undetermined numerical coefficient, which presumably 
is small at the scale set by $N$ and $Q$. Hence, for large values of 
$N$ and $Q$ the decay  is strongly suppressed. Note that in this case
the global and local vacua are not described by the same low-energy 
degrees of freedom. The processes in which small regions of the true 
vacuum (i.e., the atomic phase, that necessarily has a lower particle 
number density) could materialize within the ball, would create local  
overdensities in the ball because of the particle number conservation, 
and would be exponentially suppressed at low temperatures.

The tunneling process discussed above
describes the destruction of  the whole metanucleus. 
The metanuclei may  decay into their smaller 
counterparts via the tunneling of individual particles, or 
neutral pairs of particles escaping directly 
from the bulk of the nucleus.  We can perform estimates similar 
to the one done above  -- to escape, a particle should at least 
overcome a potential barrier with the hight of order $\sim m_H$, 
and widths of  order $\sim R_c$. Then, the tunneling rate would be 
suppressed at least by  the exponential factor, 
${\rm exp} \left (-\alpha_g Q \right )$ 
(we ignore the numerical  coefficient in the exponent).
This is strongly suppressed for large values of $Q$. \\

There are other channels through which a ball of condensate could decay.  
We start with the decay  through the evaporation of surface charges or, 
similarly, the accretion of nearby charges, if the latter are present.   
On the solution the ratio $Q/R$  is fixed, $Q/R= m_H/\alpha_g$.  
A spontaneous emission of a single charge from the surface 
would result in a new radius  $R' = (Q-1) R/Q$, with reduced 
surface energy. However, this would lead to the growth of the bulk 
Fermi degeneracy energy. To study  systematically whether the emission process
is favorable or not, we fix the ratio $Q/R$ and vary 
the energy with respect to $R$. Including the energy of the 
fermions and of the surface charge, the 
total energy (in the relativistic approximation for the fermions) is
\beq
E =\frac{3}{4}\left(\frac{9 \pi}{4}\right)^{1/3} 
\frac{N^{4/3}}{R}+ \frac{m_H^2}{\alpha_g}  R +m_HN\,.
\label{energy3}
\eeq
We have ignored the gradient of $\sigma$ in the bulk and near  
the surface, which in any event are $\lsim N/R$, 
and, hence  negligible.

Varying with respect to $R$ gives 
$R_{\rm{optimal}} \propto \alpha^{1/2}_g N^{2/3}/m_H$. 
Since emitting a charge decreases $R$, we want $R_c < R_{\rm{optimal}}$ in 
order for the condensate ball to be stable with respect to emission.  
This implies that $\alpha_g Q \ll  \alpha^{1/2}_g N^{2/3}$. 
Thus, combining all the constraints, our solution is valid as long as
\beq
1\ll  N^{1/3}\ll \alpha_g Q \ll \alpha^{1/2}_g N^{2/3} \, .  
\label{Qbound1}
\eeq
\noindent 
For $Q > \alpha^{1/2}_g N^{2/3}$ the condensate 
ball will emit charges, or decay into smaller balls (fission of metanuclei),
until $R_c = R_{\rm{optimal}}$.  Furthermore, 
when (\ref {Qbound1}) is satisfied, and 
there are other charges  present nearby, 
it is possible for the condensate ball to accrete charges, or to 
fuse with other balls,  until the  stable radius is 
reached\footnote{For the nonrelativistic fermions the condition of  
stability (\ref{Qbound1}) becomes 
$Q \lesssim \left(\frac{m_H}{m_J}\right)^{1/3} \, N^{5/9}$.}.

The above energetics arguments show that the 
electrons would not escape the surface,  as long as 
(\ref {Qbound1}) is satisfied. There should also exist 
a microscopic explanation for this, in terms of the 
local attractive forces  acting on the surface electrons.  
Perhaps, the most straightforward 
explanation would have been in terms of (quasi)localization of 
the fermions due to the bosonic background near the surface. 
This is plausible, since the scaling of the 
attractive interactions between the bulk 
condensate and the surface fermions, differs from that of  
the repulsion between  the bulk and 
surface fermions. However, a rigorous study of 
this issue would require exact solutions for the scalar and gauge fields  
within the surface layer, which are not available at present. 
There are also different mechanisms that may  
also contribute to an effective attraction  
for the surface electrons.  

The first mechanism is based on an analogy with  
an excess charge on the surface of a conductor. As it is known, up to a 
certain critical value,  this excess charge would not escape the 
surface. In terms of energetics, this can be explained as follows: if a 
single negative charge is being slightly removed from the surface,  
a positive image-charge should be put  
in the place of the removed negative one, in order to maintain 
the equipotential  surface \cite {Jackson}.  
In terms of local interactions,  
this can be understood in terms of  
the so-called double layer (dipole-like) 
structure that appears near the surface of a 
conductor  \cite {Frenkel}. Let us first consider a 
conductor with zero overall charge.  Because the 
electrons in the conductor are bound weaker than the lattice ions, 
there is a slight leakage of the electronic  charge beyond  
the surface, when the latter is defined according to the 
distribution of the lattice ions.  Let us denote 
the extent  of the leakage by $L$,  
measured as a distance from the surface.
Hence, at a distance $L$ in the exterior of the surface there is an 
excess of negative  charges $N_{-}$, and therefore,  at a distance $L$ in 
the interior,  there is an excess of positive charges $N_{+}$. 
Then, according to Gauss's law, any negative charge placed 
outside of the surface at a distance less than $L$, would experience 
a local attraction toward the surface.  This is  
what's called the double layer attraction for the  neutral 
conductors.  The above mechanism  remains approximately 
valid even when additional charge $gQ$  is  
placed on the surface, as long as $gQ \ll N_{+}$. 

Similar arguments could be presented for the surface physics of metanuclei.
The scalars and fermions,  spill out of the 
surface just a bit, to a small scale  of order 
$\Delta \sim (m_Hm_g)^{-1/2}$. It's only within this scale that the dipole
layer may form because of the difference  between the scalar 
and fermion masses and interactions.  If so, we can 
estimate the dipole charge to be 
$N_{+}\sim g J_s R_c^2 \Delta$. Then, the double-layer attraction will be 
present, as long as the  charge $N_{+}$  is much greater that the 
excess charge $gQ$. The latter condition gives $\alpha_g Q \ll 
\alpha_g^{1/5} N^{3/5}$. Typically, this would a bit stronger than  
the constraint in (\ref {Qbound1}).

As to the second possible mechanism, there may be  
a local attraction of like charges near the surface of the ball
due to relativistic effects.
Consider a single electron near the surface of 
a negatively charged ball that we're dealing with. It's 
been long known (see, e.g., \cite {Case,Zeldovich}) 
that once the magnitude of the electrostatic potential exceeds  
the electron mass,  relativistic attraction effects,
determined by minus the  potential square,
may dominate over the nonrelativistic  repulsion (we note here that this 
attraction, however, cannot  lead to quantum-mechanical bound 
states for Dirac equation).  The electrons near the surface in our 
case are in this relativistic 
regime; for instance, in one of the  example considered in the next section  
the total relevant potential  for surface electrons 
(that also includes their chemical potential)  is $\sim 5~MeV$, 
which is about 10 times greater than the electron mass. 

Whether the above two mechanisms make significant contributions
to the near-the-surface attraction, remains to be seen.

Another potential decay channel is via Schwinger pair-creation of the 
fermions (we assume that the fermions are lighter than the 
scalars in our case) or other light charged particles, 
due to the electric field near the surface of the ball.  
We consider first large size metanuclei (the ones that because of their size 
cannot have deep bound levels \cite {Zeldovich}),
for which the electric field 
$ {\cal{E}} \,= \,\frac{gQ}{4 \pi R^2}\, = 
\,\frac{4 \pi m_H^2}{g^3 Q}$  can be made subcritical by 
increasing $Q$.  The standard textbook formula for the 
pair-creation in a constant electric field may not be 
applicable here,  since the process involves  tunneling 
to infinity, in which case  the metanuclei, no matter how large, 
cannot be well-approximated  by an infinite charged plane. 
Nevertheless, one can estimate the pair-creation probability 
of particles of mass $m < m_H$  by the quasi-classical exponent, 
${\cal W}\propto  \exp({-2S})$  
\beq
S=\int_{R_0}^{\infty} |p(R)|dR, ~~~
|p(R)|\equiv \sqrt {(2m\alpha_g Q/R) -(\alpha_g^2 Q^2/R^2)}, 
\label{shw}
\eeq
where  $R_0\equiv \alpha_gQ/2m$. When the upper limit in the integral above 
is set  to infinity, corresponding to creation of a on-shell 
pair, the integral diverges,  and the probability is zero.
This is consistent with our earlier finding that the electrons 
do not escape from the surface, because of the energetics arguments. 
In more general cases, however, particles can be created with a 
nonzero energy,  which can happen when one of them ends up in a 
state of a negative energy binding with  the metanucleus,  
ensuring by its binding  the conservation of energy\footnote{Alternatively, a 
positive energy could  also be gained  if a particle of the {\it same} 
charge as the metanucleus,  attached itself to the metanucleus 
and increased  its charge and size.  
However, in this case, since the partner particle is oppositely 
charged, the electric field of the metanucleus would not 
push it out to infinity to separate it from its pair.}.

In this case, (\ref {shw}) would get 
modified. Most significantly, the upper limit of the 
integration in (\ref {shw}) would become  a finite, energy dependent, number.  
However, the exponential factor 
would still be  proportional to  $S \propto \alpha_gQ$, and the 
probability would be suppressed  for large values of $Q$. This is 
in accordance with the intuition that the electric 
field of the metanucleus decreases with increasing $Q$. 
The exponent giving the dominant contribution, when  the high energy particle 
production is allowed by the energy conservation, 
would scale as  $S \sim \alpha_gQ (m^2/m_H^2)\sim (m^2/g{\cal E})$,  
in a qualitative agreement with the Schwinger formula.

In the next section  we consider the metanuclei made of the helium-4 
ions and electrons at densities above atomic and below nuclear.
These metanuclei are of super-atomic size, and satisfy the above-discussed 
conditions of stability.  On the other hand, one could also imagine metanuclei 
made of  other scalars, such as e.g. sleptons. In this case 
the metanuclei can have a typical size of the ordinary nuclei. 
Then, the deep bound levels would be allowed, and the pair creation 
process won't be suppressed (for a review, see, \cite {Zeldovich}). 
The resulting equilibrium  object would have a shell of induced 
screening charge around it. The calculation of the distribution of 
the screening charge for that  case will be presented elsewhere.

\section{Metanuclei from electrons and helium nuclei}

The results of the previous sections can be adopted to  the 
system of  charged helium-4 nuclei $He^{++}$, and electrons $e^{-}$
(the scalar charge $g_\phi$ is twice as large as the fermion charge $g_\psi$).
We consider such a system below the nuclear  but above the atomic 
density. The former condition sets  
$Q \gg {m_H\over 200~{\rm MeV}}\frac{N^{1/3}}{\alpha_{\rm{em}}}$, 
and the latter gives  $Q \ll \frac{1}{\alpha^2_{\rm{em}}} 
\frac{m_H}{m_J} N^{1/3}$. 

Taking  $m_H \simeq 3.7 \,$GeV, $m_J \simeq 0.5\,$MeV, we find that the system 
with  $N \sim (10^{12}-10^{15})$ and $Q\sim (10^8-10^{9})$ 
satisfies all the constraints discussed in the previous sections
\footnote{There are other allowed possibilities for $N$ and $Q$. 
We choose the above numbers as typical.}.
The size of the condensate ball in this case is  
$R_c \sim (10^5 -10^6)\, {\rm{fm}}$, with the average 
inter-particle separation  $\sim (10-100) ~{\rm{fm}} $, 
the number-density of particles $\sim (2-20\, {\rm{MeV}})^3$, 
and the total energy $E_c \sim 4 \cdot (10^{12}-10^{15})\, {\rm{GeV}}$.  
These object have  a huge energy excess -- 
almost $40$ MeV per $He^{++}$ particle, in the 
simplest case.  The excess energy per particle scales 
as $\sim m_H {N^{1/3}\over \alpha_{\rm em} Q}$. There'll be huge 
energy liberated in decays of such metanuclei.

As long as temperature of the interior of the metanuclei is 
small  enough that the $He^{++}$  de Broglie 
wavelengths still overlap, the above described properties are 
expected to remain valid\footnote{For  
low enough  temperatures and densities,  
the fusion of the condensed helium nuclei, 
as well as the process of their destruction by energetic electrons, 
are expected to  be suppressed. For instance, we consider densities that 
are below the ``neutronization'' threshold of the helium nuclei.}. 
For instance, for the 
number density $\sim (10\, {\rm{MeV}})^3$, at temperatures 
below  $10^{-2}{\rm MeV}\sim 10^8 K$, 
the above described properties should be expected to hold.
The metanuclei may have formed  in starts, galaxies, or 
during some dramatic astrophysical events. 
They could represent a new state of matter, which could be 
searched for in, e.g., cosmic rays. 
One should expect, though, that their  formation and survival 
probability in the Universe to be rather low. 
Identification of the concrete mechanisms  of their 
formation in a cosmological/astrophysical 
environment, if such mechanism exist,  requires further careful 
studies of finite temperature effects, and goes beyond the 
scope of the present work.

Similar condensate balls can be ``made of'' other scalars and fermions.
Some examples are: (i) The scalars are $He^{++}$ nuclei 
and fermions are anti-protons; (ii) The scalars are condensed composite states
such as Cooper pairs, or charged pions $\pi^{\pm}$  and fermions are 
anti-protons/protons or electrons/positrons; 
(iii) In supersymmetric models the role of the 
scalars could be played by  squarks or sleptons. The helium-4 
nuclei have an advantage that they are stable states. All the particles
that can decay, such as the pions, squarks and sleptons,  should be 
captured/produced in the condensate before they could decay.

The survival probability of some of the metanuclei 
would increase if they could form neutral meta-atoms by dressing 
up with  electron/positrons. This possibility is planned to be 
discussed in future.

\vspace{0.2in}

{\it Acknowledgments.} We'd like to thank  Savas Dimopoulos, Lance Dixon, 
Misha Shifman, Matt Kleban, and especially, Andrei Gruzinov for valuable 
comments. The work  of GG is supported by 
NASA grant NNGG05GH34G. RAR is supported by  
James Arthur graduate fellowship.

\end{document}